\documentclass[12pt]{article}
\usepackage{amssymb,amsmath, epsfig}
\include{epsf}

\usepackage{graphicx}
\usepackage{amsfonts}
\usepackage{amscd}
\usepackage[all]{xy}
\advance \textheight by \topskip
\makeatletter

\renewcommand{\theequation}{\thesection.\arabic{equation}}

\newcommand{\e}{{\mathrm e}}
\newcommand{\ii}{{\mathrm i}}
\newcommand{\dd}{{\mathrm d}}
\newcommand{\eqn}[1]{(\ref{#1})}
\newcommand{\noi}{\noindent}

\def\appendix#1
{
\addtocounter{section}{1}\setcounter{equation}{0}%\setcounter{section}{1}
 \renewcommand{\thesection}{\Alph{section}}
 \section*{Appendix~\thesection\protect\indent \parbox[t]{11.715cm}{#1}}
 \addcontentsline{toc}{section}{Appendix \thesection\ \ \ #1}
}

\newcommand{\real}{{\mathbb R}} %% real numbers
\def\bra#1{\left\langle #1\right|}
\def\ket#1{\left| #1\right\rangle}
\def\hs#1#2{\left\langle #1\right.\left| #2\right\rangle}

\newcommand{\Tr}[1]{\:{\rm Tr}\,#1}
\newcommand{\be}{\begin{equation}}
\newcommand{\ee}{\end{equation}}
\newcommand{\beq}{\begin{equation}}
\newcommand{\eeq}{\end{equation}}
\newcommand{\bea}{\begin{eqnarray}}
\newcommand{\eea}{\end{eqnarray}}
\def\beqa{\begin{eqnarray}}
\def\eeqa{\end{eqnarray}}
\def\nn{\nonumber}
\newcommand{\del}{\partial}

\def\RR{{\mathbb R}}

\newcommand{\eq}{\begin{equation}}
\newcommand{\eqa}{\begin{eqnarray}}
\newcommand{\en}{\end{equation}}
\newcommand{\ena}{\end{eqnarray}}

\newcommand{\pt}{\phi^{\rm Tay}}
\newcommand{\dm}[2]{\ket{#1}\bra{#2}}

\begin{document}
\title{
\begin{flushright}
{\small CPHT-RR 123.1109}\\[-0.4cm]
{\small LPT ORSAY 09.103}\\[.51cm]
\end{flushright}
%\vskip 1cm
{\bf Curing the UV/IR mixing for field theories with translation-invariant  $\star$ products } }
\author{
{\sf   Adrian Tanasa${}^{a,b}$\thanks{e-mail:
adrian.tanasa@ens-lyon.org}
 and
{\sf Patrizia Vitale}${}^{c,d}$\thanks{e-mail:vitale@na.infn.it}}\\
{\small ${}^{a}${\it Centre de Physique Th\'eorique, CNRS UMR 7644,}} \\
{\small {\it \'Ecole Polytechnique, 91128 Palaiseau, France}}  \\
{\small ${}^{b}${\it Institutul de Fizic\u a \c si Inginerie Nuclear\u a Horia Hulubei,}} \\
{\small {\it P. O. Box MG-6, 077125 M\u agurele, Rom\^ania}}\\
{\small ${}^{c}${\it Laboratoire de Physique Th\'eorique,
Universit\'e Paris XI}} \\
{\small {\it 91405 Orsay Cedex, France}}\\
{\small ${}^{d}${\it Dipartimento di Scienze Fisiche, Universit\`{a}
di Napoli {\sl Federico II} } } \\
{\small and {\it INFN, Sezione di Napoli, Via Cintia 80126 Napoli,
Italy} } }
%\date{\today}
\maketitle

\vskip-1.5cm

\vspace{2truecm}

\begin{abstract}
\noindent The ultraviolet/infrared (UV/IR) mixing of noncommutative field
theories has been recently shown
to be a generic feature of translation- invariant associative
products. In this paper we propose to take into account the quantum
corrections of the model to modify in this way the noncommutative
action. This idea was already used to cure the UV/IR mixing for
theories on Moyal space. We show that in the present framework also,
this proposal proves successful for curing the mixing. We achieve
this task by explicit calculations of one and higher loops Feynman
amplitudes. For the sake of completeness, we compute the form of the new action in the matrix base for the Wick-Voros product.
\end{abstract}

%\vspace{2cm}
%MSC codes: \\

%Subject Classification:\\


\newpage

\section{Introduction and motivation}
\renewcommand{\theequation}{\thesection.\arabic{equation}}
\setcounter{equation}{0}
\label{introducere}

Noncommutative geometry  \cite{book-connes}  is an appealing framework for the quantization of gravity. At the Planck scale,
the quantum nature of the underlying spacetime replaces a local interaction by a specific nonlocal
effective interaction in the ordinary Minkowski space \cite{dop} .

Noncommutative quantum field theories (for general reviews, see
\cite{Szabo} and  \cite{nek}) can  be interpreted as limits 
of matrix models or of string theory models. The first use of
noncommutative geometry in string theory was in the formulation of
open string theory  \cite{witten}. Noncommutativity is here natural
just because an open string has two ends and an interaction which
involves two strings joining at their end points shares all the
formal similarities to noncommutative matrix multiplication. In this
context, one also has the Seiberg-Witten map \cite{sw}, which maps
the noncommutative vector potential to a conventional Yang-Mills
vector potential, explicitly exhibiting the equivalence between
these two classes of theories.

But probably  the simplest context in which noncommutativity  arises is in a limit in which a large background
antisymmetric tensor potential dominates the background metric. In
this limit, the world-volume theories of Dirichlet branes become
noncommutative \cite{string1}, \cite{string2}. Noncommutativity was
also recently proved to arise as some limit of loop quantum gravity
models. There, the effective dynamics of matter fields coupled to
$3-$dimensional quantum gravity is described after integration over
the gravitational degrees of freedom by some noncommutative quantum
field theory \cite{fl} .
In a different context, some $3-$dimensional noncommutative space emerging in the context of  $3-$dimensional Euclidean quantum gravity was also studied in \cite{karim}.

In condensed matter physics, noncommutative theories can be of particular interest
when describing effective non-local interactions, as is the case, for example, of the fractional
quantum Hall effect. where different authors proposed that a good description of this
phenomenon can be obtained using noncommutative rank 1 Chern-Simons theory \cite{hall1}.

Nevertheless, when going from commutative to noncommutative
theories, locality is lost and one can wonder, in this situation, if
renormalizability can be restored. Indeed, when describing theories
on the noncommutative Moyal space (the most studied noncommutative
space), a new type of non-local divergence occurs, the
UV/IR mixing \cite{melange}. This new divergence is
non-local and cannot be absorbed by counterterms at the level of the
2-point function.

Despite this important difficulty, solutions exist for
renormalizability to be restored. This is achieved  for the $\phi^4$
theory by modifying the propagation part of the initial action, such
that this new type of divergence is cured. A first type of
modification adds a harmonic oscillator term in the propagator
\cite{GW}. A different type of modification was proposed in
\cite{GMRT}, where the quantum correction $1/p^2$ was included in
the bare action in momentum space. The physical interpretation of this model in the long distance regime  was studied in \cite{helling}. Such a modified scalar propagator appears also in
recent work on non-abelian gauge theory in the context of the
Gribov-Zwanziger result \cite{qcd}. Both these noncommutative models
were proved renormalizable at any order in perturbation theory, but
the latter is also manifestly translation-invariant.  Several field
theoretical properties have been further investigated, for both
these noncommutative models (see \cite{param2}-\cite{kreimer} and
references therein). Furthermore, some algebraic geometrical
properties of the parametric representation of the Grosse-Wulkenhaar
models have been investigated in  \cite{marcolli}.

A part from the Moyal product, other noncommutative products have been investigated  to construct  noncommutative  field theories on flat space-time. One of them is the Wick-Voros (WV) product \cite{Voros}
%, \cite{BordemannWaldmann1, BordemannWaldmann2} 
which corresponds to normal ordering in  deformation  quantization \cite{Bayen}, as opposed to symmetric ordering which is instead related to the   Moyal product.
Recently, a scalar $\phi^4$ theory was studied, with this kind of noncommutativity
\cite{GLV08}.  Computing the non-planar
tadpole Feynman amplitude, it was shown that the UV/IR mixing appears in this  framework as well, although the propagator and the vertex have different forms than in the Moyal case.  This result was extended to generic translation invariant products on flat space-time \cite{GLV09}, showing in a one-loop calculation that the UV/IR mixing stays unmodified for the whole class of products.

In this paper we propose to modify the Euclidean action of scalar field theories with quartic interaction on noncommutative $\real^d$, with generalized translation-invariant star products, along the lines of \cite{GMRT}, in order to cure
 the UV/IR mixing. Namely, we compute the
one-loop quantum correction for the propagator and we modify the
scalar action accordingly. We then compute, for the modified action,
the one-loop quantum corrections for the propagator and for the
vertex and we show that the mixing has disappeared when inserting the modified non-planar tadpoles into
''bigger'' non-planar graphs. 

At tree level we show that the propagator may be decomposed  as a sum of Klein-Gordon propagators, some of which with negative sign. In a commutative setting this is a signal of illness of the theory since, when performing  a Wick rotation to the Minkowski space, these new fields  lead to  negative norm states, which in turn can be rephrased into loss of unitarity of the S matrix.   Whether or not the same conclusions can be drawn for our model is an interesting open problem, mainly  because the Minkowskian analogue of an Euclidean theory is not uniquely defined in the NC setting, but also because there is no general agreement on the  definition of particle states, commutation relations and the S matrix formalism itself (cfr \cite{ALV08, GLV08} and refs. therein). We will come back to this argument in the paper.

Quantum field theories (QFT) with the WV product (which, as already stated, are a particular class of the field theories we treat here) have already been investigated in the literature \cite{ChaichianDemichevPresnajder}.  This product has been studied within the coherent states framework \cite{HLS-J} and in relation to matrix models and Chern-Simons theory \cite{fuzzydisc}, \cite{india}.   Black holes have also been defined using such a noncommutative product \cite{indienii}.
Finally, let us also state that a different approach for studying QFT with WV product  has been undertaken in \cite{bal}. 

From a mathematical point of view, translation-invariant  $\star$ products on the (hyper-)plane are all equivalent  to the Moyal product in the sense of formal series \cite{kon} as they all share the same underlying Poisson bracket. Nevertheless, they are not a priori physically equivalent, as they yield QFTs with different quantum actions. Furthermore, one cannot relate these QFTs by simple field redefinitions, as we will show in the sequel.
\medskip

The paper is structured as follows. In the next section we recall
the definition and some basic properties of translation-invariant
generalizations of the Moyal product on Euclidean  $\real^{d}$. The results of \cite{GLV09} are recalled, showing the appearance of the UV/IR mixing.
In section $3$ we write  the action of the modified model we
propose. We also derive  the associated Feynman rules, namely the
modified propagator and modified vertex.   We also discuss the issue of ghost states in  a Minkowskian formulation.

We then compute the one-loop quantum corrections for the propagator and
the vertex of the proposed model.  This allows to show that, inserting 
non-planar tadpoles into some higher loop graphs leads to  IR convergent
Feynman amplitudes, thus curing the problem of the UV/IR mixing.  
In  section 4 we recall the matrix basis  for the WV product and we compute the expression of the modified action in this basis.
Finally, 
the last section presents some conclusions and perspectives.

\section{ Translation-invariant products}
\setcounter{equation}{0} 

In this section  we review translation-invariant star products on  the  space $\real^d$, as derived in
\cite{GLV09}. As we will see, examples of such products are the Moyal product and the less known
WV one (or normal ordered product).
 For the present purposes we present the products in terms of their integral kernel in Fourier space, although
 other forms are available. A generic star product on $\real^d$ may be represented as
\be
(\phi\star \psi)(x)=\frac1{(2\pi)^{\frac d2}}\int\dd^d p~\dd^d q ~ \dd ^d k~
\e^{\ii p \cdot x} \tilde \phi(q)\tilde \psi(k) K(p,q,k), \label{intprod}
\ee
where $K$ can be a distribution and $\tilde \phi(q)$ is the Fourier
transform of $f=\phi$. The product of $d$-vectors is understood with the
Minkowskian or Euclidean metric: $p\cdot x=p_ix^i$. The usual
pointwise product is also of this kind for
$K(p,q,k)=\delta^d(k-p+q)$. Translation invariance  requires  that the product obey
\newcommand{\tran}{\mathcal T}
\be
\tran_b(\phi)\star \tran_b(\psi)=\tran_b(\phi\star \psi),
\label{tinv0}
\ee
where  $\tran_b(f)(x)=f(x+b)$ represents the translation by the vector $b$.
At the level of Fourier transform we have
\be
\widetilde{\tran_b \phi}(q)=\e^{\ii b p}\tilde \phi(q).
\ee
It may be seen that, for  the product~\eqn{intprod} to be invariant,  the kernel must be of the form
\be
K(p,q,k)=\e^{\alpha(p,q)}\delta(k-p+q), \label{tinv}
\ee
where $\alpha$ is a generic function of $p$ and $q$, further constrained by associativity and cyclicity. 
We therefore consider products that can be expressed as
\be
(\phi\star \psi)(x)  =\frac1{(2\pi)^{\frac d2}}\int\dd^d p~ \dd^d q ~ \e^{\ii p
\cdot x} \tilde \phi(q)\tilde \psi(p-q) \e^{\alpha(p,q)}.
\label{intprodtran}
\ee
Except for  the commutative case, $d$ has to be even because of translation invariance (besides degenerate cases, where one of the dimensions commutes with the other ones). 

Let us also emphasize here that translation invariance requires as well the commutator of coordinates to be constant, as in the Moyal case. We explictly show this at the end of this section, in equation \eqref{alphacomm}.

When $\alpha=0$, one has the
usual  pointwise product.
One then has two important examples of  noncommutative associative products which are of the form above. They are both borrowed by ordinary  phase space quantization and correspond to different ordering choices. One is the
Moyal product, quite well  studied in the literature. It corresponds to symmetric ordering in deformation quantization of phase space (Weyl quantization) and in Fourier transform  it acquires the form
\be
(\phi\star_M \psi)(x)=\frac1{(2\pi)^{\frac d2}} \int \dd^{d} p \, \dd^{d} q~
\tilde \phi(q) \tilde \psi(p-q) \e^{\ii p\cdot x} \e^{\frac{\ii}{2}
p_i\theta\Sigma^{ij} q_j} \label{MoyalFourier}
\ee
thus giving
\be
 \alpha_M(p,q)=-\frac{\ii}{2}\theta \Sigma^{ij}q_ip_j. \label{almo}
\ee
We have denoted by $\Sigma$ the $d\times d $ block-diagonal antisymmetric matrix 
 \[
\Sigma = \left(\begin{array}{cccc}
0 &1 &  &  \\ -1& 0 &  & \\  &  & . & \\  &  &  & .
\end{array}\right)
\]
  and by $\theta$ some constant noncommutativity parameter. 

The other example is  the  WV product. It corresponds to normal ordering in deformation quantization of phase space  and in Fourier transform it reads
\be
(\phi\star_{WV} \psi)(x)=\frac1{(2\pi)^{\frac d2}} \int \dd^{d} p \, \dd^{d} q~
\tilde \phi(q) \tilde \psi(p-q) \e^{\ii p\cdot x} \e^{- \theta
q_-\cdot (p_+-q_+)} \label{WVFourier}
\ee
with
\be
p^i_\pm=\frac{p^i_1\pm\ii p^i_2}{\sqrt{2}} ,\ \  {i=1,..., d/2} \label{defp}
\ee
the function $\alpha$ is therefore given by
\be
\alpha_{WV}(p,q)=-\theta q_-\cdot (p_+-q_+). \label{alphavoros}\ee
Further  restrictions on $K$ come from the associativity requirement which reads
\be
\int\dd ^d k K(p,k,q)K(k,r,s)=\int\dd^d k K(p,r,k)K(k,s,q)
\label{cocyclecondition}
\ee
This is nothing but the usual cocycle condition in the Hochschild
cohomology. For more details on cohomological aspects we refer the interested reader to \cite{GLV09}.  In terms of $\alpha$, equation \eqn{cocyclecondition} reads
\be
\alpha(p,q)+\alpha(q,r)=\alpha(p,r)+\alpha(p-r,q-r)
\label{associativity}
\ee
>From this cocycle relation then follow:
\bea
\alpha(p,p)&=&\alpha(0,0)=\alpha(p,0)\nonumber\\
\alpha(0,p)&=&\alpha(0,-p)\nonumber\\
\alpha(p,q)&=&-\alpha(q,p)+\alpha(0,q-p)
 \label{relations}
\eea
and
\be
\alpha(p+q,p)+\alpha(-p-q,-q).+\alpha(0,p+q)-\alpha(0,p)-\alpha(0,q)=0.
\label{useful}
\ee
 The second relation of \eqref{relations} ensures also the trace
property. We have indeed
\bea
\int\dd^d x~\phi\star  \psi&=&\int\dd^d x~ \dd^d p ~\dd^d q ~\e^{\alpha(p,q)} \e^{\ii p\cdot x} \tilde \phi(q) \tilde \psi (p-q)\nonumber\\
&=&\int \dd^d q~ \e^{\alpha(0,q)} \tilde \phi(q) \tilde \psi(-q)= \int \dd^d x~ \psi \star \phi .
\eea
%We also require the algebra to be a $*$-algebra, namely there
%is a $*$ conjugation such that ${\phi^*}^*=\phi$ and $(\phi\star
%\psi)^*=\psi^*\star \phi^*$. This latter relation imposes
%\be
%\alpha(p,q)^*=\alpha(-p,q-p) \label{starcondition}
%\ee
For the product to be commutative $\alpha$ has to satisfy the condition
\be
\alpha(p,q)=\alpha(p,p-q),
 \label{commutativeprod}
\ee
which may be regarded as a coboundary condition (see again \cite{GLV09}). Finally, let us  notice that the Moyal and WV products are related by the following relation:
\be
\alpha_M(p,q)=\alpha_{WV}(p,q)-\frac{\theta}{2}q\cdot(p-q).
\ee

\medskip

As already stated above, translation-invariant products  have been introduced in \cite{GLV09} in the context of  noncommutative scalar field theories with quartic interaction, on $\real^d$ with Minkowskian metric. Here we choose to work with Euclidean metric. We keep the notation $d$ for the number of space-time dimensions; nevertheless, when explicit divergence analysis is performed, we refer to the case
$$ d=4.$$
 The action reads
\be
S=\int \dd^d x~ \left[\frac{1}{2}(-\del_\mu \phi \star\del_\mu \phi +m^2 \phi\star\phi) +\frac{\lambda}{4!}
\phi\star\phi\star\phi\star\phi\right].
\ee
In momentum space the propagator is thus
\begin{equation}\label{TPGF}
{G}^{(2)}_0(p)=\frac{e^{-\alpha(0,p)}}{p^{2}+m^{2}} 
\end{equation}
whereas for the vertex we have
\begin{equation}\label{V}
V_{\star}=V_0
e^{\alpha(k_{1}+k_{2},k_{1})+\alpha(k_{3}+k_{4},k_{3})+\alpha(0,k_{1}+k_{2})},
\end{equation}
and  we have denoted by $V_0$ the ordinary commutative  vertex
\be
V_0= \frac{\lambda}{4!} (2\pi)^d\delta^d\left(\sum_{a=1}^4
{k_a}\right).    \label{V0def}
\ee
Interestingly, a propagator of the form \eqn{TPGF} with the function $\alpha$ specified   in \eqn{alphavoros} was already found in \cite{smailagicspallucci} in a different approach.

To obtain the  four-point correlation  function at
tree level we just attach to the vertex four
propagators. We thus have (up to a constant):
\be
{G}_0^{(4)}=\frac{e^{\alpha(k_{1}+k_{2},k_{1})+\alpha(k_{3}+k_{4},k_{3})+\alpha(0,k_{1}+k_{2})-\sum_{a=1}^{4}\alpha(0,k_{a})}}
{\prod_{a=1}^{4}(k_{a}^{2}+m^{2})}\delta\left(\sum_{a=1}^{4}k_{a}\right). 
\label{g4tree}
\ee

\medskip

Let us now make the following important remark. In order to obtain the usual Feynman rules of a Moyal QFT, one can try  by reabsorbing the phase $e^{-\alpha(0,p)}$ of \eqref{TPGF} by a proper field redefinition. Nevertheless, this would not reproduce the vertex form of Moyal QFT. We thus conclude that, by a simple field redefinition one does not have equivalence between the general class of QFTs we deal with in this paper and the Moyal one. In \cite{GLV08} it was shown that an equivalence between Moyal and WV Minkowskian QFTs can be established at the level of the S-matrix only by implementing the appropriate twisted Poincar\'e symmetry for each of  them.

\medskip

Before going further we also give some explanations on the planarity of the Feynman graphs used in this work. As in the Moyal case, the vertex has a symmetry under cyclic permutation of the incoming/outgoing fields at some vertex. Furthermore, one can also use some matrix base to re-expresses these products (for the WV case, see section $4$). For all these reasons, an appropriate way to represent Feynman graphs is through ribbon graphs. If the genus of the manifold on which the respective graph is denoted is vanishing, the respective graph is planar. Furthermore, another important notion is the one of faces broken by external legs. In the rest of the paper, by a slight abuse of language we will call non-planar graphs also the planar graphs with more than one face broken by external legs.

\medskip

\begin{figure}[htb]
\epsfxsize=2.5 in
\bigskip
\centerline{\epsffile{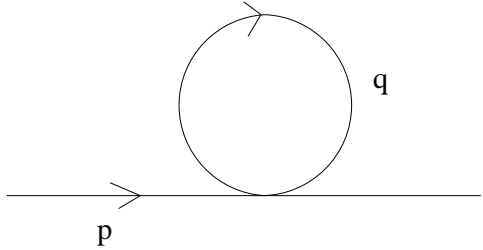}}
\caption{\sl The planar tadpole graph.}
\label{tad}
\end{figure}

\begin{figure}[htb]
\epsfxsize=2.5 in
\bigskip
\centerline{\epsffile{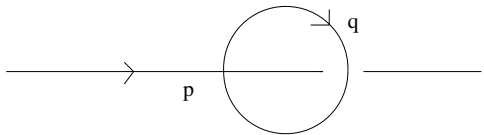}}
\caption{\sl The non-planar tadpole graph.}
\label{tadnp}
\end{figure}

Consider now the two graphs of Figures \ref{tad} and \ref{tadnp}.

For the planar case of Figure \ref{tad} the correction is obtained using three propagators of the form \eqref{TPGF},
one with momentum $p$, one with momentum $-p$, one with momentum
$q$ and the vertex \eqref{V} with assignments $k_{1}=-k_{4}=p$ and $k_{2}=-k_{3}=q$
and, of course, the integration in $q$. We have (up to a constant)
\be
G^{(2)}_{1,\, P}=\frac{\e^{-\alpha(0,p)}}{(p^{2}+m^{2})^{2}} \int\dd^d q\,\frac{\e^{\sigma(p,q)}}
{ (q^{2}+m^{2})} \label{planar}
\ee
where we have used the fact that the three exponential factors of
the vertex combine with two out of three exponential factors of the
propagators to yield
\be
\sigma(p,q)= \alpha(p+q,p)+\alpha(-p-q,-q)+\alpha(0,p+q)-\alpha(0,q)-\alpha(0,p). \label{sigmapq}
\ee
Using the cocycle condition \eqn{useful},  we have then 
\be
\sigma (p,q)=0
\ee
 Notice that, with respect to
the commutative case, the only correction is in the factor
$\e^{-\alpha(0,p)}$ which is the correction of the free propagator.
The ultraviolet divergences here are thus identical to the commutative and to the Moyal case.
%and therefore
%the short distance physics is unaffected (in this aspect) by the
%star product.

Consider now the non-planar case in Figure~\ref{tadnp}.
The structure is the same as in the planar case, but this time the assignments are
\begin{equation}
k_{1}=-k_{3}=p\quad\mathrm{and}\quad k_{2}=-k_{4}=q.
\end{equation}
We have (up to a constant)
\be
G^{(2)}_{1,\,  NP}=\int\dd^d q\,\frac{\e^{-\alpha(0,p)+\alpha(p+q,p)-\alpha(p+q,q)} \e^{\sigma(p,q)}}
{(p^{2}+m^{2})^{2}(q^{2}+m^{2})}. 
\label{nonplanardiagram}
\ee
with the same notation as above. Therefore $\sigma(p,q)=0$ and 
 the  one-loop corrections to the propagator in the non-planar case
can be rewritten as
\begin{equation}
G^{(2)}_{1, NP}=\frac{\e^{-\alpha(0,p)}} {(p^{2}+m^{2})^{2}},\int\dd^d q\,\frac{\e^{\omega(p,q)}} {(q^{2}+m^{2})}, \label{nptp0}
\end{equation}
where we have introduced the antisymmetric function
\begin{equation}
\omega(p,q)=\alpha(p+q,p)-\alpha(p+q,q). \label{omega}
\end{equation}
For the 
Moyal product this term is the oscillating phase.
For general translation invariant products this   function has been computed explicitly
in \cite{GLV09}, using the properties of the function $\alpha$. It does not depend on the
specific translation invariant product but only on the cohomology class of $\alpha$.
Except for numerical factors, one has, for all translation-invariant products
\be
\omega(p,q)  =\ii p_i\theta\Sigma^{ij}q_j =2 \alpha_M(p,q).
\label{omega2}
\ee
Details on the derivation may be found in \cite{GLV09}. The function
$\alpha(0,p)$  which appears in \eqn{nptp0} is not integrated in the
loop, therefore it does not influence the convergence properties of
the graphs.

In fact the function $ \omega$ is nothing but the Poisson structure of the underlying classical space, the germ of
deformation
of the commutative product towards the star product. It  determines the noncommutativity of space-time coordinates.
A straightforward calculation gives
\be
x^i\star x^j-x^j\star x^i=-
\frac{\partial^{2}\alpha}{\partial p_{i}\partial q_{j}}(0,0)+
\frac{\partial^{2}\alpha}{\partial p_{j}\partial q_{i}}(0,0)= \omega^{ij} = i\theta\Sigma^{ij}.
\label{alphacomm}
\ee
%Therefore we have shown that products with the same Poisson
%tructure (and hence the same commutator) which are  equivalent in
%the sense of Kontsevitch, have the same structure of
%infrared/ultraviolet mixing. Indeed, u
Using \eqref{omega2}, one can
prove that the Feynman integral \eqref{nptp0} is UV finite but has a
\beqa
\label{behaviour}
\frac{C_1}{(\theta  p)^2}+m^2 C_2 {\rm log} (\theta\ p)^2 + F(p)
\eeqa
behavior in the IR regime of the external momentum $p$. We have denoted by $C_1$ and $C_2$ some constants
and by $F(p)$ some analytic function at $p=0$ (see \cite{GMRT} for a detailed analysis).

\section{The proposed model and quantum corrections - curing the UV/IR mixing}
\renewcommand{\theequation}{\thesection.\arabic{equation}}
\setcounter{equation}{0}

In this section we write  the action for the proposed model in the noncommutative setting described  previously. We then compute quantum corrections of the modified model  (one and higher number of loops) which  show the way in which the UV/IR is manifestly cured.

As already stated in the Introduction, the modification we propose for the model of the previous section is dictated by the quantum corrections. Thus, we add to the action the  supplementary term:
\beqa
\label{supco} \delta S[\phi]=\frac{a}{2\theta^2} \int \dd^{d}x \,  (\del_\mu) ^{-1}\phi \star( \del_\mu) ^{-1} \phi.
\eeqa
The complete action, in coordinate space, reads then
\be
S=\int \dd^d x~ \left[\frac{1}{2}(\del_\mu \phi \star\del_\mu \phi  +\frac{a}{\theta^2} \del_\mu^{-1}\phi \star \del_\mu^{-1} \phi +m^2 \phi\star\phi) +\frac{\lambda}{4!} \phi\star\phi\star\phi\star\phi\right] ,
 \label{moda}
\ee
with
\be
\del_\mu ^{-1}\phi (x) = \int \dd^\mu x'~ \phi(x'). \label{intop}
\ee
The supplementary term is   better understood in momentum space. Observing that for the star product \eqn{intprodtran}
\be
\int \dd^d x \, f(x)\star g(x) = \int \dd^d p \, \tilde f(p) \tilde g(-p) \e^{\alpha(0,p)}
\ee
we have
\be
\delta S[\phi]=\frac{a}{2\theta^2} \int \dd^{d}p \,   \frac{\e^{\alpha(0,p)}}{p^2} \tilde \phi(p)\tilde \phi(-p). \label{sup}
\ee
This leads to a modification of the propagator \eqref{TPGF} in the form
\be
\label{TPGF-nou}
{G}^{(2)}_0(p)=\frac{e^{-\alpha(0,p)}}{p^{2}+m^{2}+\frac{a}{(\theta p)^2}} \ .
\ee
Note that, as in \cite{GMRT} the new parameter $a$ is taken to be positive (such that the propagator \eqref{TPGF-nou} is positively defined. The vertex contribution remains the same as in \eqref{V} (since the term \eqref{sup} is only quadratic in the field).

A few remarks are to be done here. The first one is related to the possibility of decomposing the propagator \eqref{TPGF-nou} as a sum of conventional Klein-Gordon propagators. We use the formula
\be
\frac{1}{A+B}=\frac 1A - \frac 1A B \frac{1}{A+B},
\ee
for 
\be
A=p^2+m^2,\ \ B=\frac{a}{\theta^2 p^2}.\label{AB}
\ee
Thus, the propagator \eqref{TPGF-nou} writes
\begin{eqnarray}
\label{dec}
&&\frac{e^{-\alpha(0,p)}}{p^2+m^2}-\frac{e^{-\alpha(0,p)}}{p^2+m^2}\frac{a}{\theta^2 p^2 (p^2+m^2)+a}\cr
&&=
\frac{e^{-\alpha(0,p)}}{p^2+m^2}-\frac{e^{-\alpha(0,p)}}{p^2+m^2}\frac{a}{\theta^2 (p^2 +m_1^2)(p^2+m^2_2)},
\label{propa2}
\end{eqnarray}
where $-m_1^2$ and $-m_2^2$ are the roots of the denominator of 
the second term in the LHS considered as a second order equation in $p^2$, namely
\begin{eqnarray}
m_1^2&=& \frac{\theta^2 m^2- \sqrt{\theta^4 m^4 - 4 \theta^2 a}}{2\theta^2}\nn\\
m_2^2&=& \frac{\theta^2 m^2+\sqrt{\theta^4 m^4 - 4 \theta^2 a}}{2\theta^2}
\end{eqnarray}
with $0<a<\theta^2 m^4/4$. Also
\be
m^2>m_2^2>\frac{m^2}{2}>m_1^2>0
\ee
 Note that this decomposition was already made (for the Moyal case) in \cite{Geloun:2008hr}. We now go further 
and decompose the second term on the RHS of \eqref{dec}; after some algebra this finally  leads to rewrite the propagator \eqn{TPGF-nou} as  an alternate sum 
\be
{G}^{(2)}_0(p)=\frac{a/\theta^2}{m_2^2-m_1^2}\left(\frac{1}{m_1^2(p^2+m_2^2)}-\frac{1}{ m_2^2(p^2+m_1^2)}\right)
\label{decom}
\ee
In  similar situations, in the commutative framework  (for example   QFT with higher derivatives) an equivalent description is introduced  in terms of KG fields. The ones responsible for the negative propagators are the ghost fields. They lead,  when analitically continuing the model to the Minkowskian setting, to states of negative norm. They are not independent fields, therefore one could resort to the original formulation in terms of one field (for scalar theories as the one considered here)  but this usually causes the loss of uninitarity for the  S matrix. Some care is needed in order to extend this analysis to  our model.
The first main difference between commutative and noncommutative field theory is that the Minkowskian version of a NC field theory is not uniquely defined. Important features, such as the UV/IR mixing of the Euclidean formulation may be completely absent in some Minkowskian formulations \cite{bahns}..
The second important point is that the very concept of particle state is not well defined in a nonlocal theory. This remark is particularly important for our model, where the asymptotic regime is not attained for small noncommutativity  \cite{helling}.
However, let us stick to the usual Wick rotation. Then, it was shown in \cite{unitarity}  that time-space noncommutativity leads to perturbative nonunitarity, independently from the details of the model, although a more careful analysis indicates that unitarity can be restored \cite{liao} . Later on it has been argued with a nonperturbative study  \cite{CLZ02}  that unitarity loss  is a direct consequence of the UV/IR mixing. Whether this problem persists in our model, where the modification to the kinetic term was introduced precisely to cancel the UV/IR mixing, it is a delicate issue and we don't know the answer.  In order to see the consequences of the ghost fields appearing in \eqn{decom} in the appropiate Minkowskian formulation,  one should carefully define the S matrix and then study its properties. This is an interesting problem which  deserves further investigation.

Let us also make a second remark with respect to the Euclidean theory treated here. Even though, for massive theories, the quantum correction of the propagator \eqref{behaviour} has sub-leading logarithmic divergence, one does not need to take it into consideration when proposing a model which may be {\it perturbatively} renormalizable. We will come back to this point at the end of the next section.

Since we have modified the propagator \eqn{TPGF} to \eqn{TPGF-nou}
but not the vertex  \eqn{V},
the one-loop corrections to the propagator represented by the graphs  of Figures \ref{tad} and \ref{tadnp} may be derived as in \eqn{planar}, \eqn{nonplanardiagram} and we find respectively:
\be
G^{(2)}_{1,\, P}=\frac{\e^{-\alpha(0,p)}}{(p^{2}+m^{2}+\frac{a}{\theta^2p^2} )^{2}} \int\dd^d q\,\frac{\e^{\sigma(p,q)}}
{ (q^{2}+m^{2} +\frac{a}{\theta^2q^2})},
\ee
\begin{equation}
G^{(2)}_{1,\, NP}=\frac{\e^{-\alpha(0,p)}}
{(p^{2}+m^{2} +\frac{a}{\theta^2p^2})^{2}}\int\dd^d q\,\frac{\e^{\omega(p,q)+\sigma(p,q)}}
{(q^{2}+m^{2}) +\frac{a}{\theta^2q^2}}
\label{nptp}
\end{equation}
with $\sigma(p,q)=0$.
 
\begin{figure}[htb]
\epsfxsize=3.0 in
\bigskip
\centerline{\epsffile{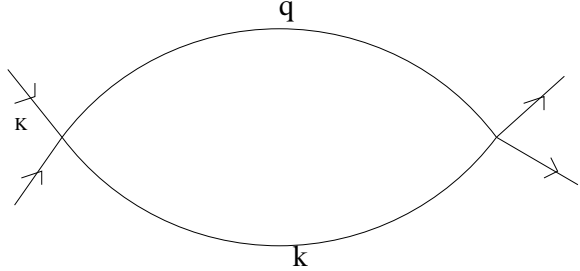}}
\caption{\sl Planar one loop four point graph of incoming momentum $K$}
\label{planarfish}
\end{figure}

Let us consider now the one-loop correction to $4$-point functions.  For the planar case in Fig.~\ref{planarfish}
we have
\bea
G^{(4)}_{1,\, P}&=&\frac{\lambda^2}{8}G_0^{(4)} \int\dd^d
q\, \dd^d k\,\frac{\delta(K-q-k)\e^{\sigma(q,k)}}
{(q^2+m^2+ \frac{a}{\theta^2 q^2})(k^2+ m^2+ \frac{a}{\theta^2 k^2} )},
\nonumber\\\label{G4planar}
\eea
with  $ G_0^4 $ the  4-points Schwinger function at tree level, given by \eqn{g4tree}, while $
\sigma(q,k)=0$
 (with $\sigma$ defined in \eqn{sigmapq}).
Note that we have also denoted by $K$ the total incoming momentum.

\begin{figure}[htb]
\epsfxsize=3.0 in
\bigskip
\centerline{\epsffile{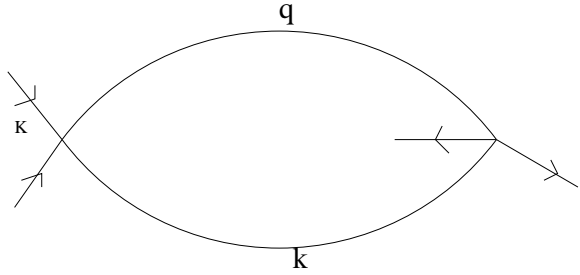}}
\caption{\sl A non-planar one loop four points graph of total incoming momentum $K$}
\label{nonplanarfish}
\end{figure}

For the non-planar case, one possible graph is shown in  Fig. \ref{nonplanarfish}.
We have then  
\bea
\label{log}
 G^{(4)}_{1,\, NP}&=&\frac{\lambda^2}{8}\tilde G_0^4\int\dd^d q\,\dd^d k
 \frac{\delta(K-q-k)\e^{\omega(q,k)+\sigma(k,q)}}
{(q^2+m^2+ \frac{a}{\theta^2 q^2})(k^2+ m^2+ \frac{a}{\theta^2 k^2} )},
\label{G4nonplanar}
\eea
with $\omega(q,k)$ defined in \eqn{omega}, \eqn{omega2} while the remaining exponential factors rearrange in such a way to yield   $\sigma(k,q)$ vanishing. The other non-planar graphs are obtained similarly, with a relabeling  of the external paths.

For $d=4$, the last integral behaves like a logarithm in the external momenta.  Indeed, this can be seen from the fact that, in the UV regime of the loop momentum $q$, the corrections in $a/(\theta^2q^2)$ are neglectable.  Solving the $\delta$ function, the integral \eqref{log} behaves like
\beqa
\int d^4 q \frac{e^{\omega(q,k)}}{q^4}.
\eeqa
Because of the form \eqref{omega2} of the factor $\omega(q,k)$ one can easily obtain the logarithm behaviour in the external momenta. Nevertheless, these logarithms are harmless for {\it perturbative} renormalization. We will come back on this point at the end of this section.

\begin{figure}[htb]
\epsfxsize=4.0 in
\bigskip
\centerline{\epsffile{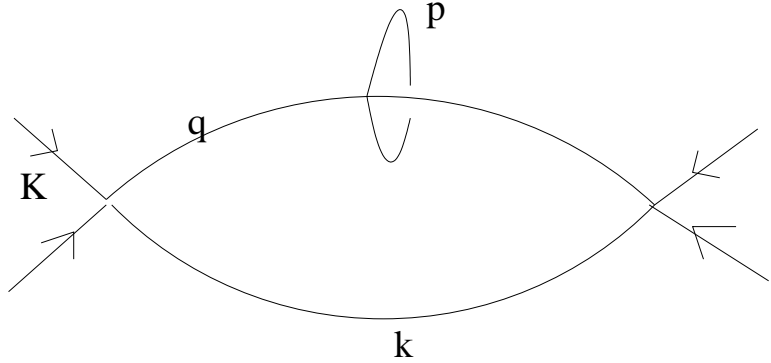}}
\caption{\sl A $2-$loop graph obtained by inserting into the bubble graph of Fig. \ref{planarfish} a non-planar tadpole of momentum $p$.}
\label{prima}
\end{figure}

\begin{figure}[htb]
\epsfxsize=4.0 in
\bigskip
\centerline{\epsffile{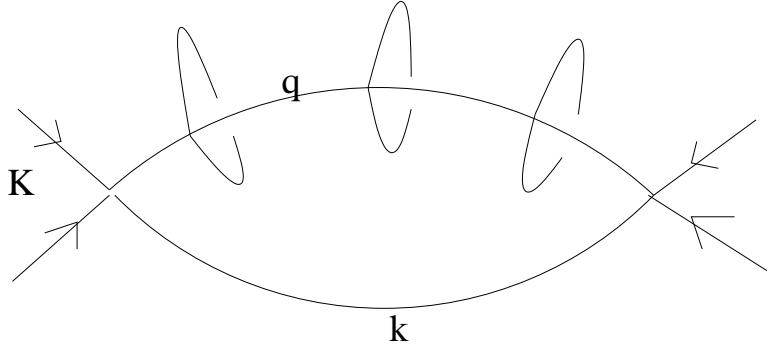}}
\caption{\sl A  non-planar graph obtained from the insertion into the bubble graph of Fig. \ref{planarfish} of a chain of  non-planar tadpoles of momenta $p_i$ ($i=1,\ldots,N$).}
\label{N}
\end{figure}

Let us now compute some  $2-$loop contributions to the  $4-$point Schwinger function. The first of them is the one  due to  the graph in Fig \ref{prima}. 
This graph is
obtained by inserting the non-planar tadpole \eqn{nptp} of momentum $p$  into the planar graph in fig \ref{planarfish}.
One has
\be
\label{1tad}
   G_{2, \, P}^{(4)}= 
G_0^{(4)}\int\dd^4 p \, \frac{\e^{\omega(p,q)+\sigma(p,q)}}{p^2+m^2+
\frac{a}{\theta^2 p^2}} \int\dd^d q \, \dd^4 k\, 
\frac{\delta(K-q-k)\e^{\sigma(q,k)}}
{(q^2+m^2+ \frac{a}{\theta^2 q^2})^2(k^2+ m^2+ \frac{a}{\theta^2
k^2})},
\ee
and $\sigma=0$. 
Let us emphasize that again, the cancellation of the exponentials  in $\sigma$ is obtained thanks to the cocycle condition \eqref{useful}. It is this cancellation that makes that the Feynman integrals have the same behaviour as in the Moyal case.
%We also compute a
%$2-$loop contribution to the non-planar $4-$points Schwinger function
% in Fig
%??. 
%which is obtained by inserting the non-planar tadpole of momentum $p$ into the non-planar graph in
%Fig \ref{nonplanarfish}.  The only difference with the previous case is an extra factor $\e^{\omega(q,k)}$ which plays no %r\^ole in the UV/IR issue (this type of factor acting as some kind of cut-off in the UV regime of the respective momentum).
%One has:
%\be
%\delta_{(\mathrm {2-loop})}  \tilde G_{NP}^{(4)}= \tilde G_0^4\int
%\dd^d p \, \frac{\e^{\omega(p,q)}}{p^2+m^2+ \frac{a}{\theta^2
%p^2}}\int\dd^d q \, \frac{\delta(k_1+k_2-q-k)\e^{\omega(q,k)}}
%{(q^2+m^2+ \frac{a}{\theta^2 q^2})^2(k^2+ m^2+ \frac{a}{\theta^2
%k^2} )}
%\ee
Let us now investigate the behaviour of the Feynman amplitude of the more general graph obtained from inserting a chain of $N$ non-planar tadpoles in the planar graph of Fig. \ref{planarfish} (see Fig. \ref{N}).

%As before, the $\alpha$ factors cancel out by repeatedly  applying  the cocycle condition.   
The integral to investigate writes indeed
\beqa
\label{unu}
\int \prod_{i=1}^Nd^dp_i\frac{e^{\omega(p_i,q)+\sigma(p_i,q)}}{p_i^2+m^2+\frac{a}{\theta^2p_i^2}}\int \dd^d q\,\dd^d k\, (\frac{1}{q^2+m^2+\frac{a}{\theta^2q^2}})^{N+1}
\frac{\delta(K-q-k) \e^{\sigma(q,k)}}{k^2+m^2+\frac{a}{\theta^2q^2}}
\eeqa
with $\sigma(p_i,q)=\sigma(q,k)=0$. 
This is the generalization of the Feynman amplitude \eqref{1tad}. Let us now have a closer look at the structure of the divergences of this general integral. For $d=4$, when performing the integrations in the momenta $p_i$ ($i=1,\ldots,N$) and placing ourselves in the IR regime of the momentum $q$, each of these integrals leads to a $1/{\theta^2 q^2}$ behaviour (as proved above). The integral \eqref{unu} thus becomes
\beqa
\label{doi}
\int d^4 q 
(\frac{1}{\theta^2q^2})^N
(\frac{1}{q^2+m^2+\frac{a}{\theta^2q^2}})^{N+1}
\frac{1}{(q-K)^2+m^2+\frac{a}{\theta^2(q-K)^2}}
\eeqa
Note that  if $a=0$ this integral is IR divergent for $N>1$ (for $N=1$  the mass $m$ prevents the divergence to appear). Nevertheless, if $a\ne 0$,  in the IR regime of $q$ the dominant term is the $a/{\theta^2q^2}$ in the propagators and the integral
% behaves like
%\beqa
%(\theta^2 q^2)
%\eeqa
leads to an IR finite behaviour.

\medskip

\begin{figure}
\epsfxsize=5.0 in
\centerline{\epsfig{figure=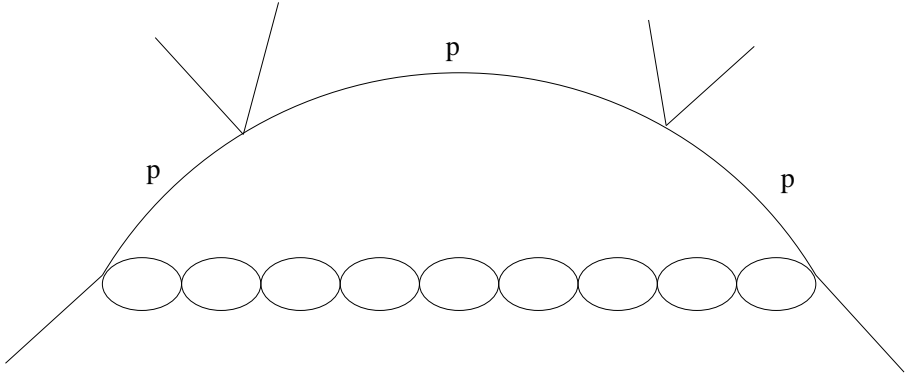,width=12cm}}
%\hfil \epsfig{figure=figraz-8.ps,width=2cm}}
\caption{Insertion of a chain of bubble graph into some bigger graph.}
\label{renormaloni}
\end{figure}

Let us now discuss  the appearance of some logarithmic divergences
at the level of the $2-$ and $4-$point functions, logarithms on the
external momenta of the respective graph. This divergences are
harmless when dealing with {\it perturbative} renormalizability. To
illustrate this, let us consider the example of Fig.
\ref{renormaloni}, where one has to deal with a chain of $N$
non-planar bubble graphs inserted into some ``bigger''
non-planar graph. Since, as already explained above, the correction
proposed here in the propagator is irrelevant in the UV regime
(where this analysis is now performed), the Feynman integral gives
the same behaviour as in a commutative theory, namely:
\beqa
\int \dd^4 p ~\frac{1}{(p^2+m^2)^3}{\rm log}^N \frac{p^2}{m^2}
\approx N!
\eeqa
This is a (large) finite number which appears as a difficulty in
summing perturbation theory - the renormalon problem (see for example 
\cite{book-rivasseau}). The situation is analogous for the
non-planar tadpoles insertions. These logarithms should however
be taken into consideration when defining a model which is requested
to be non-perturbatively renormalizable. Let us also remark that in
\cite{GMRT}, these logarithm divergences have not appeared because
of the use of some appropriate scale decomposition (the multi-scale
analysis being used there for the proof of perturbative
renormalization).

\section{The modified action for the  WV product in the matrix basis}
\renewcommand{\theequation}{\thesection.\arabic{equation}}
\setcounter{equation}{0}

For the WV product it is possible to rewrite the model in a
suitably defined  matrix basis. The basis is a variation of the one described in \cite{bondiavarilly} for the Moyal product and it  was   introduced, up to our knowledge,  in
\cite{fuzzydisc} where the WV product was used  on $\real^2$
to build a fuzzy version of the disc.  In the following we review
the derivation of the matrix basis as in  \cite{fuzzydisc} and we adapt
it to the present notation;  we then derive the  model under
consideration in the matrix basis. It is convenient to consider the
plane as a complex space with $z=(x+iy)/\sqrt{2}$. The quantized
versions of $z$ and $\bar z$ are the usual annihilation and creation
operators, $a=(\hat x +i \hat y)/\sqrt{2 } $ and $a^\dagger=(\hat x
-i \hat y)/\sqrt{2}$ with a slightly unusual normalization, so that
their commutation rule is
\be
[a,a^\dagger]=\theta. 
\label{aadcomm}
\ee
 Given the function $\phi(\bar z,z)$ consider its Taylor
expansion:
\be
\phi(\bar z, z)=\sum_{m,n=0}^\infty \pt_{mn}\bar z^m z^n. \label{taylorphi}
\ee
To this function we associate the operator
\be
\Omega_\theta(\phi):=\hat\phi=\sum_{m,n=0}^\infty\pt_{mn}{a^\dagger}^m
a^n. \label{taylorexp}
\ee
We have thus ``quantized'' the plane using a normal ordering
prescription.  The map $\Omega_\theta$ is invertible. It can be efficiently expressed defining the \emph{coherent}
states:
\be
a\ket{z}=z\ket{z},
\ee
One then has
\be
\Omega^{-1}_\theta(\hat\phi)=\phi(\bar z,z)=\bra{z}\hat
\phi\ket{z}. \label{Omegam1}
\ee
The maps $\Omega$ and $\Omega^{-1}$ yield a procedure of going
back and forth from functions to operators. Moreover,  the product
of operators being noncommutative,  a noncommutative $\star$ product
between functions is implicitly defined as
\be
\left(\phi\star\phi'\right)(\bar z,
z)=\Omega^{-1}\left(\Omega(\phi)\, \Omega(\phi')\right)  .
\label{starint}
\ee
It is possible to see that the WV product \eqn{WVFourier} is exactly of this form, namely
\be
(\phi\star_{WV} \psi) (\bar z, z)= \bra{z}\hat \phi \hat
\psi\ket{z}.\label{WVpr}
\ee
  There is another useful basis on which it is possible
to represent the operators and hence the functions - 
the matrix basis. As we already stated above,  it is  similar  to the one
introduced for the Moyal product in \cite{bondiavarilly} and subsequently
used in \cite{GW}.  Consider the number operator
\be
{\rm N}=a^\dagger a , \label{defnumb}
\ee
and its eigenvectors which we indicate
by $\ket{n}$:
\be
{\rm N}\ket{n}=n\theta\ket{n}  .
\ee
We can then express the operators within a density matrix notation:
\be
\hat\phi=\sum_{m,n=0}^\infty\phi_{mn}\dm{m}{n}  .
\label{dmexp}
\ee
Applying the dequantization map  \eqn{Omegam1} to $\hat \phi$,  we
associate to  it
\be
\phi(z,\bar z)= \bra{z}\phi\ket{z}= \sum_{m,n=0}^\infty\phi_{mn}
\hs{z}{m}\hs{n}{z}:=\sum_{m,n=0}^\infty\phi_{mn} v_{mn}(z,\bar z)
\label{phimat}
\ee
to be compared with  \eqn{taylorphi}, the same function  in a different basis. Observing that
\be
\hs{z}{n}=\e^{-\frac{\bar z z}{2\theta} } \frac{\bar z} {\sqrt{n!\theta^n}}
\ee
we have
\be
v_{nm}(\bar z, z)=\e^{-\frac{\bar z z}{2\theta} }\frac{ \bar z^n z^m
}{\sqrt{n!m! \theta^{n+m}}} \label{matrixbasis}.
\ee
The elements of the density matrix basis have a very simple multiplication
rule
\be
\ket{m}\hs{n}{p}\bra{q}=\delta_{np}\dm{m}{q}; \label{densmult}
\ee
this leads to
\be
v_{mn}(\bar z, z)\star_{WV} v_{pq}(\bar z, z)=\delta_{np} v_{mq}(\bar z,
z).
\ee
Moreover, one has:
\be
\int \dd^2 z~ v_{nm}(\bar z, z) = \delta_{nm} \theta\pi.
\label{norm}
\ee
The  functions $ v_{nm}(\bar z, z)$ thus form an orthogonal
basis in the noncommutative algebra of functions on the plane, with
the WV product, the matrix basis (in analogy with
its operator counterpart, \eqn{densmult}). The connection between the
expansions~\eqn{taylorexp} and~\eqn{dmexp} is given by:
\bea
a&=&\sum_{n=0}^\infty \sqrt{(n+1)\,\theta}\dm{n}{n+1}\label{a}\\
a^\dagger&=&\sum_{n=0}^\infty
\sqrt{(n+1)\,\theta}\dm{n+1}{n}\label{adag}.
\eea
Thus, looking at their symbols
\bea
z&=&\bra{z}a\ket{z}= \sum_n \sqrt{(n+1)\theta} \; v_{nn+1}(\bar z, z) \nn\\
\bar z&=&\bra{z}a^\dag\ket{z}= \sum_n \sqrt{(n+1)\theta} \; v_{n+1
n}(\bar z, z),
\eea
we then have
\be
\pt_{mn}= \sum_{l=0}^{min\{m,n\}} (-1)^l \frac{\phi_{m-l,n-l}}{l ! \sqrt{(m-l)!(n-l)!\theta^{m+n}}} . \label{ptp}
\ee
In the density matrix basis, using ~\eqn{matrixbasis}, the
product  \eqn{WVFourier} (or  \eqn{WVpr}) simplifies to an infinite row by column matrix
multiplication:
\be
\left(\phi\star_{WV}\psi\right)_{mn}=\sum_{k=1}^\infty \phi_{mk}\psi_{kn}.
\ee
Using the expansion ~\eqn{phimat} and \eqn{norm} it is easy to see
that 
\be
\int \dd^2z\, \phi(\bar z,z)={\pi\theta}
\Tr\Phi={\pi\theta}\sum_{n=0}^\infty\phi_{nn}, \label{defint}
\ee
where we have introduced the infinite matrix $\Phi$ with entries
$\{\phi_{ij}\}$.

Let us generalize this basis to higher (even) dimensions. In $d$
dimensions we need $d/2$ copies of the WV plane, with $d/2$
pairs of complex coordinates $z^i,\bar z^i$. As for the Moyal
hyperplane, coordinates describing different 2-planes, commute among
themselves. We define
\be
v_{\vec m \vec n}(\bar z_1,...,\bar z_{d/2}, z_1,...,z_{d/2})=
\hs{z_1,...z_{d/2}}{m_1,...,m_{d/2}}\hs{n_1,...,n_{d/2}}{z_1,...z_{d/2}}=\Pi_{i=1}^{d/2}
v^i_{m_i n_i}
\ee
with
\be
v^i_{m_i n_i}=\hs{z_i}{m_i}\hs{n_i}{z_i}=\e^{-\frac{\bar z_i
z_i}{2\theta} }\frac{ \bar z_i^{m_i} z_i^{n_i} }{\sqrt{n_i!m_i!
\theta^{n_i+m_i}}}
\ee
The functions $v_{\vec m \vec n}$ form an orthogonal  basis in the
algebra of noncommutative functions on the hyperplane, as it may be
easily checked that
\be
v_{\vec m \vec n}\star_{WV} v_{\vec p \vec q} = v_{\vec m \vec q}
\delta_{\vec n \vec p}.\label{vprodd}
\ee
Moreover, one has
\be
\int\dd^{d}z\; v_{\vec m \vec n}= (\pi\theta)^{d/2} \delta_{\vec m \vec
n}. \label{normd}
\ee
 The field $\phi$ is therefore expanded as
\be
\phi(\bar z,
z)=\sum_{\stackrel{m_i,n_i=0}{i=1,...,d}}^\infty\phi_{\vec m \vec n}
v_{\vec m \vec n}(z,\bar z).\label{phi2d}
\ee
Moreover, one has
\be
\phi(\bar z, z)\star_{WV}\psi(\bar z, z)= \sum \phi_{\vec m \vec
n}\psi_{\vec n \vec q} v_{\vec m \vec q}
\ee
and also 
\be
\int \dd^{d} z~ \phi(\bar z,z)=(\pi\theta)^d\sum\phi_{\vec m \vec
n} \delta_{\vec m \vec n}.
\ee
 Having established this 
basis it is now possible to express the action of our model
\eqn{moda} in matrix notation. The mass and the interaction terms
are just row by column multiplication. We have
\bea
&&\frac{m^2}{2}\int \dd^{d} z ~\phi\star_{WV} \phi
+\frac{\lambda}{4!}\int \dd^{d} z~ \phi\star_{WV}
\phi\star_{WV}\phi\star_{WV}\phi\nn\\
&&=(\pi\theta)^{d/2}\left(\frac{\mu^2}{2}\sum
\phi_{\vec m \vec n}\phi_{\vec n \vec q} \delta_{\vec m \vec
q}
+ \frac{\lambda}{4!}\sum\phi_{\vec m \vec n}\phi_{\vec n \vec
q}\phi_{\vec q \vec p}\phi_{\vec p \vec r}\delta_{\vec m \vec
r}\right)\label{mass+int}
\eea
Let us consider the kinetic term. We first observe that
\be
\frac{1}{2}\int \dd^{d}z
\,\del_\mu\phi\star_{WV}\del_\mu\phi=\int\dd^{{d}}z\, \sum_i
\del_{z_i}\phi\star_{WV}\del_{\bar z_i}\phi\label{kin}
\ee
and
\bea
\del_{z_i}\phi&=&\frac{1}{\theta}[\bar z_i,\phi]_{\star_{WV}}\\
\del_{\bar z_i}\phi&=&\frac{1}{\theta}[ z_i,\phi]_{\star_{WV}}.
\eea
Eqs. \eqn{a}, \eqn{adag}, suitably generalized to the $d$ case,
imply in turn that $z_i$ and $\bar z_i$ are expanded in the matrix
basis as
\bea
z_i&=&\sum_{\vec n}\sqrt{{(n_i+1)}{\theta}}\, v^i_ {n_i \,n_i+1}\Pi_{j\ne i}v^j_{n_j \,n_j}\\
\bar z_i&=&\sum_{\vec n} \sqrt{{(n_i+1)}{\theta}}\, v^i_ {n_i+1
\,n_i} \Pi_{j\ne i}v^j_{n_j \,n_j}.
\eea
We then have
\bea
\del_{z_i}\phi&=&\frac{1}{\theta} \sum_{\vec m\vec n}\phi_{\vec
m\vec n} \left(\sqrt{m_i+1}v^i_{m_i+1\, n_i}-\sqrt{n_i}
v^i_{m_i \,n_i-1}\right)\Pi_{j\ne i }v^i_{m_i\, n_i}\nn\\
\del_{\bar z_i}\phi&=&\frac{1}{\theta} \sum_{\vec m\vec n}\phi_{\vec
m\vec n} \left(\sqrt{m_i}v^i_{m_i-1\, n_i}-\sqrt{n_i+1} v^i_{m_i\,
n_i+1}\right)\Pi_{j\ne i }v^i_{m_i \,n_i}
\eea
The kinetic term \eqn{kin} becomes 
\bea
&&\frac{(\pi\theta)^d}{\sqrt \theta} \sum_{\stackrel{\vec m \vec n}{ \vec p \vec
q}}\phi_{\vec m\vec n}\phi_{\vec p\vec q}\sum_i \bigl(\delta_{ m_i
q_i} \delta_{n_i p_i}(-m_i-n_i-1)\bigr. \nn\\
&&\left.+ \delta_{ m_i+1 q_i}\delta_{n_i\, p_i-1}\sqrt{q_i
p_i}+\delta_{ n_i-1\, p_i}\delta_{ m_i q_i+1}\sqrt{(q_i+1)
(p_i+1)}\right)\Pi_{j\ne i} \delta_{m_jq_j}\delta_{n_jp_j}
\eea

\medskip

Let us consider now the supplementary term we have added  in \eqn{moda}
\be
\frac{1}{\theta^{d}}\int \dd^{d}z\,  \left(\int
\dd^{d}z'\,\phi(\bar z',z')\right)\star_{WV} \left(\int \dd^{d}z'\,\phi(\bar z',z')\right). \label{intpro}
\ee
To compute the two indefinite integrals above we use the Taylor expansion \eqn{taylorexp}.  We thus arrive at the following expression:
\be
\frac{1}{\theta^{d}}\sum_{\stackrel{\vec m\vec n}{\vec r\vec s} }\pt_{\vec m\vec n}\pt_{\vec r\vec s} \Pi_i\frac{1}{(m_i+1)(n_i+1)(r_i+1)(s_i+1)}\int  \dd^{d}z\,  \bar z_i^{m+1} z_i^{n+1}\star_{WV} \bar z_i^{r+1} z_i^{s+1}
\ee
We then use the  matrix basis to evaluate the star product  and we obtain:
\bea
&&\frac{1}{\theta^{d}}\sum_{\stackrel{\vec m\vec n}{\vec r\vec s} }\pt_{\vec m\vec n}\pt_{\vec r\vec s}  \sum_{\vec p}  \int \dd^{d}z\, \Pi_i \frac{\bar z_i^{p_i+m_i} z_i^{p_i+n_i-r_i+s_i}\e^{-\frac{\bar z_i z_i}{\theta}}}{(m_i+1)(n_i+1)(r_i+1)(s_i+1)} \nn\\  &&\frac{(p_i+n_i+1)!\sqrt{(p_i+m_i+1)!(p_i+n_i-r_i+s_i+1)! \theta^{m_i+n_i+r_i+s_i+4}  } } {p_i!(p_i+n_i-r_i)!\sqrt{(p_i+m_i)!(p_i+n_i-r_i+s_i)!\theta^{2p_i+m_i+n_i-r_i+s_i}}}.
\eea
The integral may be easily performed and, after some algebra, we obtain for the supplementary term the expression:
\be
\sum_{\stackrel{\vec m\vec n}{\vec r\vec s} }\pt_{\vec m\vec n}\pt_{\vec r\vec s}  \sum_{\vec p} \Pi_i \frac{(p_i+m_i+1)!(p_i+n_i+1)!\theta^{r_i-m_i-p_i}}{(m_i+1)(n_i+1)(r_i+1)(s_i+1) p_i!(p_i+n_i-r_i)!}  \delta_{m_i+r_i,n_i+s_i}.
\ee
Finally, using \eqn{ptp} to rewrite the coefficients $\pt$, we get
%we arrive at the following:
\bea
&&  \sum_ {\stackrel {\vec m\vec n}{\vec r\vec s}  \vec p} \Pi_i \frac{(p_i+m_i+1)!(p_i+n_i+1)!\theta^{r_i-m_i-p_i} }{(m_i+1)(n_i+1)(r_i+1)(s_i+1) p_i!(p_i+n_i-r_i)!}  \delta_{m_i+r_i,n_i+s_i}\nn\\
&& \sum_{\vec l=0}^{min \{\vec m,\vec n\}}  \sum_{\vec k=0}^{min\{\vec r,\vec s\}} \Pi_j\frac{(-1)^{l_j+k_j} \theta^{-m_j-r_j}\phi_{\vec m-\vec l \vec n-\vec l}\phi_{\vec r-\vec k \vec s-\vec k } }{l_j!k_j!\sqrt{(m_j-l_j)!(n_j-l_j)!(r_j-k_j)!(s_j-k_j)! }}.
\label{suppmat}
\eea
where, with an abuse of notation, $ \sum_{\vec l=0}^{min \{\vec m,\vec n\}} $ stands for $ \sum_{ \stackrel{l_i=0}{i=1,..,d}}^{min \{ m_i, n_i\}}$.
Finally, summing all the contributions that we obtain from \eqn{mass+int}, \eqn{kin} and \eqn{suppmat}, the complete action \eqn{moda} of our model with supplementary term, is rewritten in the matrix basis as
\bea
S[\phi]&=&
(\pi\theta)^{d/2}\left(\frac{m^2}{2}\sum
\phi_{\vec m \vec n}\phi_{\vec n \vec q} \delta_{\vec m \vec
q}+\frac{\lambda}{4!}\sum\phi_{\vec m \vec n}\phi_{\vec n \vec
q}\phi_{\vec q \vec p}\phi_{\vec p \vec r}\delta_{\vec m \vec
r}\right)\nn\\
&+&\frac{(\pi\theta)^{d/2}}{\sqrt \theta} \sum_{\stackrel{\vec m \vec n}{ \vec p \vec
q}}\phi_{\vec m\vec n}\phi_{\vec p\vec q}\sum_i \bigl(\delta_{ m_i
q_i} \delta_{n_i p_i}(-m_i-n_i-1)\bigr. \nn\\
&&\left.+ \delta_{ m_i+1 q_i}\delta_{n_i\, p_i-1}\sqrt{q_i
p_i}+\delta_{ n_i-1\, p_i}\delta_{ m_i q_i+1}\sqrt{(q_i+1)
(p_i+1)}\right)\Pi_{j\ne i} \delta_{m_jq_j}\delta_{n_jp_j}\nn\\
&+& \sum_ {\stackrel {\vec m\vec n}{\vec r\vec s}  \vec p} \Pi_i \frac{(p_i+m_i+1)!(p_i+n_i+1)!\theta^{r_i-m_i-p_i} }{(m_i+1)(n_i+1)(r_i+1)(s_i+1) p_i!(p_i+n_i-r_i)!}  \delta_{m_i+r_i,n_i+s_i}\nn\\
&& \sum_{\vec l=0}^{min \{\vec m,\vec n\}}  \sum_{\vec k=0}^{min\{\vec r,\vec s\}} \Pi_j\frac{(-1)^{l_j+k_j} \theta^{-m_j-r_j}\phi_{\vec m-\vec l \vec n-\vec l}\phi_{\vec r-\vec k \vec s-\vec k } }{l_j!k_j!\sqrt{(m_j-l_j)!(n_j-l_j)!(r_j-k_j)!(s_j-k_j)! }}.
\eea

\section{Conclusion and perspectives}
\renewcommand{\theequation}{\thesection.\arabic{equation}}
\setcounter{equation}{0}

In this paper we have proposed a solution for curing the UV/IR mixing which appears when implementing field theories using a translation-invariant product on $\RR^4$. This solution generalizes the one proposes in \cite{GMRT} for curing the UV/IR mixing on Moyal space. We explicitly compute the $1-$loop Feynman amplitudes of the proposed model, as well as higher loops amplitudes of some graphs obtained by an insertion of  non-planar tadpoles. Our result is mainly due to the cocycle condition \eqref{relations}.

An immediate perspective is to obtain a proof of the perturbative renormalization of the proposed model at any order in perturbation theory. This could be achieved by investigating the general form of the factor generalizing the Moyal oscillating phase of some Feynman amplitude. 
%This would represent a generalization to our framework of the work of \cite{filk}, where this was done for Moyal scalar field theory.

As already stated in this paper, the noncommutative products that we have worked here are equivalent in the formal series sense of Kontsevich. We thus explicitly show that an important field theoretical result - curing the UV/IR mixing - can be obtained applying the same recipe as in Moyal field theory. It is interesting to further understand the explicit relation between this formal series equivalence of the noncommutative products and the equivalence of the Euclidean field theories thus implemented.

Nevertheless, as already argued in \cite{GLV08}, when doing Minkowskian field theory, the situation is manifestly different, because the different factors on the external propagator can lead to a different $S$ matrix form, unless 
properly implementing the quantum symmetry of the model as a twisted symmetry.
%using in an appropriate way a new concept of a twisted $S$ matrix.

\bigskip

\noi
{\bf Acknowledgment:}   Adrian Tanasa 
acknowledges 
the 
%was partially supported by the 
CNCSIS grant ``Idei'' 454/2009, ID-44. Patrizia Vitale acknowledges 
the European Science
Foundation Exchange Grant 2595 under the Research Networking Program ``Quantum Geometry and Quantum Gravity'' .
The authors would also like to warmly thank LPT Orsay for the hospitality. 
%was partially supported by the Research Networking Programme of the European Science Foundation exchange grant 


\begin{thebibliography}{99}
\bibitem{book-connes}
Alain Connes and Matilde Marcolli, ''Noncommutative Geometry, Quantum Fields and Motives''.

\bibitem{dop}
S.~Doplicher, K.~Fredenhagen and J.~E.~Roberts,
``The Quantum structure of space-time at the Planck scale and quantum
fields,''
  Commun.\ Math.\ Phys.\  {\bf 172}, 187 (1995)
  [arXiv:hep-th/0303037].
  %%CITATION = CMPHA,172,187;%%

\bibitem{Szabo}
R.~J.~Szabo,
  ``Quantum field theory on noncommutative spaces,''
  Phys.\ Rept.\  {\bf 378}, 207 (2003)
  [arXiv:hep-th/0109162].
  %%CITATION = PRPLC,378,207;%%

\bibitem{nek}
M.~R.~Douglas and N.~A.~Nekrasov,
``Noncommutative field theory,''
  Rev.\ Mod.\ Phys.\  {\bf 73}, 977 (2001)
  [arXiv:hep-th/0106048].
  %%CITATION = RMPHA,73,977;%%


\bibitem{witten}
 E.~Witten,
  ``Noncommutative Geometry And String Field Theory,''
  Nucl.\ Phys.\  B {\bf 268}, 253 (1986).
  %%CITATION = NUPHA,B268,253;%%

\bibitem{sw}
 N.~Seiberg and E.~Witten,
  ``String theory and noncommutative geometry,''
  JHEP {\bf 9909}, 032 (1999)
  [arXiv:hep-th/9908142].
  %%CITATION = JHEPA,9909,032;%%



\bibitem{string1}
Connes A, Douglas M. R., Schwarz A.:
''Noncommutative Geometry and Matrix Theory: Compactification on Tori'',
JHEP 9802, 3-43 (1998)
%hep-th/9711162

%\cite{Douglas:1997fm}
\bibitem{string2}
  M.~R.~Douglas and C.~M.~Hull,
 ``D-branes and the noncommutative torus,''
  JHEP {\bf 9802}, 008 (1998)
  [arXiv:hep-th/9711165].
  %%CITATION = JHEPA,9802,008;%%

%\cite{Freidel:2005me}
\bibitem{fl}
  L.~Freidel and E.~R.~Livine,
``Effective 3d quantum gravity and non-commutative quantum field theory,''
  Phys.\ Rev.\ Lett.\  {\bf 96}, 221301 (2006)
  [arXiv:hep-th/0512113].
  %%CITATION = PRLTA,96,221301;%%

%\cite{Joung:2008mr}
\bibitem{karim}
  E.~Joung, J.~Mourad and K.~Noui,
``Three Dimensional Quantum Geometry and Deformed Poincare Symmetry,''
  J.\ Math.\ Phys.\  {\bf 50}, 052503 (2009)
  [arXiv:0806.4121 [hep-th]].
  %%CITATION = JMAPA,50,052503;%%



\bibitem{hall1}
Susskind L.: The Quantum Hall Fluid and Non-Commutative Chern Simons Theory.
hep-th/0101029;
%\bibitem{hall2}
Polychronakos A. P.:
Quantum Hall states on the cylinder as unitary matrix Chern-Simons theory.
JHEP, { 06}, 70-95 (2001);
%\bibitem{hall3}
Hellerman S., Van Raamsdonk M.:
Quantum Hall physics equals noncommutative field theory.
JHEP { 10}, 39-51 (2001).

%\cite{Minwalla:1999px}
\bibitem{melange}
  S.~Minwalla, M.~Van Raamsdonk and N.~Seiberg,
  ``Noncommutative perturbative dynamics,''
  JHEP {\bf 0002}, 020 (2000)
  [arXiv:hep-th/9912072].
  %%CITATION = JHEPA,0002,020;%


%\bibitem{ls}
%E.~Langmann and R.~J.~Szabo,
%  ``Duality in scalar field theory on noncommutative phase spaces,''
%  Phys.\ Lett.\  B {\bf 533}, 168 (2002)
%  [arXiv:hep-th/0202039].
  %%CITATION = PHLTA,B533,168;%%


%\bibitem{GW1}
%Grosse H. and Wulkenhaar R.,
%Power-counting theorem for non-local matrix models and renormalization,
% Commun.\ Math.\ Phys. {254}, 91-127 (2005)
%[arXiv:hep-th/0305066]


\bibitem{GW}
  H.~Grosse and R.~Wulkenhaar,
  ``Renormalisation of phi**4 theory on noncommutative R**2 in the matrix base,''
  JHEP {\bf 0312}, 019 (2003)
  [arXiv:hep-th/0307017].
  %%CITATION = JHEPA,0312,019;%%\\
Renormalizationof $\phi^4$-theory on noncommutative ${\mathbb R}^4$ in the matrix
base,
 Commun.\ Math.\ Phys. { 256}, 305-374 (2005)
%[arXiv:hep-th/0401128]

%\cite{Gurau:2008vd}
\bibitem{GMRT}
  R.~Gurau, J.~Magnen, V.~Rivasseau and A.~Tanasa,
  ``A translation-invariant renormalizable non-commutative scalar model,''
Commun. Math. Phys.   {\bf 287}, 275 (2009)
  [arXiv:0802.0791 [math-ph]].
  %%CITATION = CMPHA,287,275;%%
  %%CITATION = ARXIV:0802.0791;%%
%\cite{Galluccio:2008wk}



\bibitem{helling} R.~ C. Helling,  J. You, ``Macroscopic Screening of Coulomb Potentials From UV/IR-Mixing'',
JHEP {\bf 0806}, 067 (2008)
[arXiv:0707.1885 [hep-th]].
 %%CITATION = ARXIV:07.1885,%%
%\cite{Dudal:2008sp}
\bibitem{qcd}
  D.~Dudal, J.~A.~Gracey, S.~P.~Sorella, N.~Vandersickel and H.~Verschelde,
  %``A refinement of the Gribov-Zwanziger approach in the Landau gauge: infrared
  %propagators in harmony with the lattice results,''
  Phys.\ Rev.\  D {\bf 78}, 065047 (2008)
  [arXiv:0806.4348 [hep-th]].
  %%CITATION = PHRVA,D78,065047;%%



%\bibitem{GW5}
% H.~Grosse and R.~Wulkenhaar,
%  ``Renormalisation of phi**4-theory on non-commutative R**4 to all orders,''
%  Lett.\ Math.\ Phys.\  {\bf 71}, 13 (2005).
  %%CITATION = LMPHD,71,13;%%


%\bibitem{param1}
%R.~Gur\u au and V.~Rivasseau,
%  `Parametric representation of noncommutative field theory,''
%  Commun.\ Math.\ Phys.\  {\bf 272}, 811 (2007)
%  [arXiv:math-ph/0606030].
  %%CITATION = CMPHA,272,811;%%

%\cite{Rivasseau:2007qx}
\bibitem{param2}
V.~Rivasseau and A.~Tanasa,
``Parametric representation of 'covariant' noncommutative QFT models,''
arXiv:math-ph/0701034, Commun. Math. Phys. {\bf 279}, 355 (2008)
  %%CITATION = MATH-PH/0701034;%%

%\cite{Gurau:2007az}
\bibitem{mellin}
 R.~Gurau, A.~P.~C.~Malbouisson, V.~Rivasseau and A.~Tanasa,
 ``Non-Commutative Complete Mellin Representation for Feynman Amplitudes,''   Lett.\ Math.\ Phys.\  {\bf 81}, 161 (2007)
 arXiv:0705.3437 [math-ph].
  %%CITATION = ARXIV:0705.3437;%%

\bibitem{dimreg}
R.~Gurau and A.~Tanasa,
 ``Dimensional regularization and renormalization of non-commutative QFT,''
Annales Henri Poincare, {\bf 9}, 655 (2008)
arXiv:0706.1147 [math-ph].
  %%CITATION = ARXIV:0706.1147;%%


\bibitem{hopf}
A.~Tanasa and F.~Vignes-Tourneret,
 ``Hopf algebra of non-commutative field theory,''
J. Noncomm. Geom. {\bf 2} ,
arXiv:0707.4143 [math-ph].
  %%CITATION = ARXIV:0707.4143;%%


%\cite{deGoursac:2007uv}
\bibitem{goldstone}
A.~de Goursac, A.~Tanasa and J.~C.~Wallet,
``Vacuum configurations for renormalizable non-commutative scalar models,'' Eur.\ Phys.\ J.\  C {\bf 53}, 459 (2008)
arXiv:0709.3950 [hep-th].
  %%CITATION = ARXIV:0709.3950;%%

%\cite{Tanasa:2007xq}
\bibitem{Tanasa:2007xq}
  A.~Tanasa,
  ``Feynman amplitudes in renormalizable non-commutative quantum field
  theory,'' sollicited by Modern Encyclopedia Mathematical Physics
  arXiv:0711.3355 [math-ph].
  %%CITATION = ARXIV:0711.3355;%%


\bibitem{param}
A. Tanasa, ``Overview of the parametric representation of renormalizable
non-commutative field theory'', J. Phys. Conf. Series  {\bf 103}, 012012 (2008)
 [arXiv:0709.2270 [hep-th]].
  %%CITATION = 00462,103,012012;%%


%\cite{Geloun:2008hr}
\bibitem{Geloun:2008hr}
  J.~B.~Geloun and A.~Tanasa,
``One-loop $\beta$ functions of a translation-invariant renormalizable
noncommutative scalar model,''
  Lett.\ Math.\ Phys.\  {\bf 86}, 19 (2008)
  [arXiv:0806.3886 [math-ph]].
  %%CITATION = LMPHD,86,19;%%

%\cite{Tanasa:2008bt}
\bibitem{Tanasa:2008bt}
  A.~Tanasa,
``Parametric representation of a translation-invariant renormalizable
noncommutative model,''
  J.\ Phys.\ A  {\bf 42}, 365208 (2009)
  [arXiv:0807.2779 [math-ph]].
  %%CITATION = JPAGB,A42,365208;%%

%\cite{Magnen:2008pd}
\bibitem{Magnen:2008pd}
  J.~Magnen, V.~Rivasseau and A.~Tanasa,
``Commutative limit of a renormalizable noncommutative model,''
  Europhys.\ Lett.\  {\bf 86}, 11001 (2009)
  [arXiv:0807.4093 [hep-th]].
  %%CITATION = EULEE,86,11001;%%


%\cite{Tanasa:2008gg}
\bibitem{Tanasa:2008gg}
  A.~Tanasa,
  %``Scalar and gauge translation-invariant noncommutative models,''
  Rom.\ J.\ Phys.\  {\bf 53}, 1207 (2008)
  [arXiv:0808.3703 [hep-th]].
  %%CITATION = RRPQA,53,1207;%%

%\cite{Krajewski:2008fa}
\bibitem{Krajewski:2008fa}
  T.~Krajewski, V.~Rivasseau, A.~Tanasa and Z.~Wang,
  ``Topological Graph Polynomials and Quantum Field Theory, Part I: Heat Kernel
  Theories,''
  arXiv:0811.0186 [math-ph].
  %%CITATION = ARXIV:0811.0186;%%


%\cite{Tanasa:2009hb}
\bibitem{kreimer}
  A.~Tanasa and D.~Kreimer,
``Combinatorial Dyson-Schwinger equations in noncommutative field theory,''
  arXiv:0907.2182 [hep-th].
  %%CITATION = ARXIV:0907.2182;%%

%\cite{Aluffi:2008sy}
\bibitem{marcolli}
  P.~Aluffi and M.~Marcolli,
``Feynman motives of banana graphs,''
  arXiv:0807.1690 [hep-th].
  %%CITATION = ARXIV:0807.1690;%%



\bibitem{Voros} A. Voros, ``Wentzel-Kramers-Brillouin method in the
    Bargmann representation'', Phys.\ Rev.\ A40, 6814 (1989).
  %%CITATION = PHRVA,A40,6814;%%
%\bibitem{BordemannWaldmann1}
  M.~Bordemann and S.~Waldmann,
  ``A Fedosov Star Product of Wick Type for K\"ahler Manifolds,''
  Lett.\ Math.\ Phys.\ 41, 243 (1997).
  arXiv:q-alg/9605012.
  %%CITATION = Q-ALG/9605012;%%
%\bibitem{BordemannWaldmann2} 
M.~Bordemann and S.~Waldmann,
  ``Formal GNS Construction and States in Deformation
  Quantization,'' Comm.\ Math.\ Phys.\ 195, 549 (1998).
  arXiv:q-alg/9607019.
  %%CITATION = Q-ALG/9607019;%%

\bibitem{Bayen} F.\ Bayen, in Group Theoretical Methods in Physics,
    ed. E. Beiglbock , et. al. [Lect. Notes Phys. 94, 260 (1979)];


\bibitem{GLV08}
  S.~Galluccio, F.~Lizzi and P.~Vitale,
``Twisted Noncommutative Field Theory with the Wick-Voros and Moyal
Products,''
  Phys.\ Rev.\  D {\bf 78}, 085007 (2008)
  [arXiv:0810.2095 [hep-th]].
  %%CITATION = PHRVA,D78,085007;%%

%\cite{Galluccio:2009ss}
\bibitem{GLV09}
  S.~Galluccio, F.~Lizzi and P.~Vitale,
``Translation Invariance, Commutation Relations and Ultraviolet/Infrared
Mixing,''
  JHEP {\bf 0909}, 054 (2009)
  [arXiv:0907.3640 [hep-th]].
  %%CITATION = JHEPA,0909,054;%%

\bibitem{ALV08} 
 P.~Aschieri, F.~Lizzi and P.~Vitale,
  %``Twisting all the way: from Classical Mechanics to Quantum Fields,''
  Phys.\ Rev.\  D {\bf 77}, 025037 (2008)
  [arXiv:0708.3002 [hep-th]].
  %%CITATION = PHRVA,D77,025037;%%

\bibitem{bahns}
 D.~Bahns,
  ``The ultraviolet-finite Hamiltonian approach on the noncommutative
  Minkowski space,''
  Fortsch.\ Phys.\  {\bf 52}, 458 (2004)
  [arXiv:hep-th/0401219].
D.~Bahns, S.~Doplicher, K.~Fredenhagen and G.~Piacitelli,
  ``Ultraviolet finite quantum field theory on quantum spacetime,''
  Commun.\ Math.\ Phys.\  {\bf 237}, 221 (2003)
  [arXiv:hep-th/0301100].

\bibitem{unitarity} J. Gomis and T. Mehen, “Space-time noncommutative field theories and unitarity,”
Nucl. Phys. {\bf B 591}, 265  (2000)  [arXiv:hep-th/0005129]. ~~
 L. Alvarez-Gaume, J. L. Barbon and R. Zwicky, “Remarks on time-space noncommutative
field theories,” JHEP {\bf 0105}, 057 (2001)  [arXiv:hep-th/0103069].~~
 D. Bahns, S. Doplicher, K. Fredenhagen, G. Piacitelli,  ''On the unitarity problem in space/time noncommutative
 theories``, Phys. Lett. {\bf B 533}, 178 (2002)

\bibitem{liao}
  Y.~Liao and K.~Sibold,
  ``Time-ordered perturbation theory on noncommutative spacetime. II.
  Unitarity,''
  Eur.\ Phys.\ J.\  C {\bf 25}, 479 (2002)
  [arXiv:hep-th/0206011].
  %%CITATION = EPHJA,C25,479;%% 

\bibitem{CLZ02} C. S. Chu, J. Lukierski and W. J. Zakrzewski, “Hermitian analyticity,
IR/UV mixing and unitarity of noncommutative field theories,”  	Nucl.Phys. {\bf B 632}, 219  (2002)  [arXiv:hep-th/
0201144].
\bibitem{smailagicspallucci} 
  A.~Smailagic and E.~Spallucci,
  ``Feynman path integral on the noncommutative plane,''
  J.\ Phys.\ A  {\bf 36}, L467 (2003)
  [arXiv:hep-th/0307217].
  %%CITATION = JPAGB,A36,L467;%%

\bibitem{ChaichianDemichevPresnajder} M.~Chaichian, A.~Demichev and
    P.~Presnajder,
  ``Quantum field theory on noncommutative space-times and the persistence  of
  ultraviolet divergences,''
  Nucl.\ Phys.\  B {\bf 567} (2000) 360
  [arXiv:hep-th/9812180].
  %%CITATION = NUPHA,B567,360;%%


\bibitem{HLS-J} A.~B.~Hammou, M.~Lagraa and
    M.~M.~Sheikh-Jabbari,
  ``Coherent state induced star-product on R(lambda)**3 and the fuzzy
  sphere,''
  Phys.\ Rev.\  D {\bf 66}, 025025 (2002)
  [arXiv:hep-th/0110291].
  %%CITATION = PHRVA,D66,025025;%%
%\bibitem{pinzulstern} 
A. Pinzul and  A. Stern, 
{\it A new class of two-dimensional noncommutative spaces}, JHEP
{\bf 0203}, 039 (2002) [arXiv:hep-th/0112220];\\
%%CITATION = HEP-TH 0112220;%%\\
{\it Absence of the holographic principle in noncommutative
Chern-Simons  theory}, JHEP {\bf 0111}, 023 (2001)
[arXiv:hep-th/0107179].
%%CITATION = HEP-TH 0107179;%%\\

G.~Alexanian, A.~Pinzul and A.~Stern, {\it Generalized Coherent
State Approach to Star Products and Applications to the Fuzzy
Sphere}, Nucl.\ Phys.\  {\bf B 600}, 531 (2001)
[arXiv:hep-th/0010187].
%%CITATION = HEP-TH 0010187;%%
%\bibitem{daoud}
M.~Daoud,
``Extended Voros product in the coherent states framework,''
  Phys.\ Lett.\  A {\bf 309}, 167 (2003).
  %%CITATION = PHLTA,A309,167;%%



\bibitem{fuzzydisc}
%\cite{Lizzi:2005zx}
  F.~Lizzi, P.~Vitale and A.~Zampini,
``The fuzzy disc,''
  JHEP {\bf 0308}, 057 (2003)
  [arXiv:hep-th/0306247].
  %%CITATION = JHEPA,0308,057;%%
\\
 ``The beat of a fuzzy drum: fuzzy Bessel functions for the disc,''
  JHEP {\bf 0509}, 080 (2005)
  [arXiv:hep-th/0506008].
  %%CITATION = JHEPA,0509,080;%%

\bibitem{india} 
A.P.~Balachandran, K,~Gupta and
S.~K\"urk\c{c}\"{u}o\v{g}lu, {\it Edge Currents in Noncommutative
Chern--Simons Theory from a new Matrix Model}, Syracuse Preprint
SU-4252-783, Saha Preprint SINP/TNP/03-19, [arXiv:hep-th/0306255].
%%CITATION = HEP-TH 0306255;%%

%\cite{Banerjee:2009xx}
\bibitem{indienii}
  R.~Banerjee, S.~Gangopadhyay and S.~K.~Modak,
``Voros product, Noncommutative Schwarzschild Black Hole and Corrected Area
Law,''
  arXiv:0911.2123 [hep-th].
  %%CITATION = ARXIV:0911.2123;%%

\bibitem{bal}
A.~P.~Balachandran and M.~Martone,
  ``Twisted Quantum Fields on Moyal and Wick-Voros Planes are Inequivalent,''
  Mod.\ Phys.\ Lett.\  A {\bf 24}, 1721 (2009)
  [arXiv:0902.1247 [hep-th]]; 
A.~P.~Balachandran, A.~Ibort, G.~Marmo and M.~Martone,
``Inequivalence of QFT's on Noncommutative Spacetimes: Moyal versus
Wick-Voros,''
  arXiv:0910.4779 [hep-th].

\bibitem{kon} 
M.~Kontsevich,
  ``Deformation quantization of Poisson manifolds, I,''
  Lett.\ Math.\ Phys.\  {\bf 66} (2003) 157
  [arXiv:q-alg/9709040].
  %%CITATION = LMPHD,66,157;%%

%\bibitem{filk}
%  T.~Filk,
%  ``Divergencies in a field theory on quantum space,''
%  Phys.\ Lett.\  B {\bf 376} (1996) 53.
  %%CITATION = PHLTA,B376,53;%%




\bibitem{book-rivasseau}
V. Rivasseau, ''From perturbative to constructive renormalization``, Princeton University Press, 1991.
\bibitem{bondiavarilly} J. M. Gracia-Bondia and J. C. Varilly, “Algebras of Distributions Suitable For Phase
Space Quantum Mechanics. 1,” J. Math. Phys. {\bf 29} (1988) 869.
\end{thebibliography}
\end{document}